\documentclass{webofc}
\usepackage[varg]{txfonts}   
\usepackage{listings}
\usepackage{hyperref}
\begin{document}
\title{Cepheids with the eyes of photometric space telescopes}
%
%

\author{\firstname{L\'{a}szl\'o} \lastname{Moln\'ar}\inst{1}\fnsep\thanks{\href{mailto:molnar.laszlo@csfk.mta.hu}{\tt molnar.laszlo@csfk.mta.hu}} \and
        \firstname{Andr\'as} \lastname{P\'al}\inst{1,2} \and  \firstname{Emese} \lastname{Plachy}\inst{1} 
}

\institute{Konkoly Observatory, MTA CSFK, Konkoly Thege Mikl\'os \'ut 15-17, H-1121 Budapest, Hungary  
          \and
              E\"otv\"os Lor\'and University, H-1117 P\'azm\'any P\'eter s\'et\'any 1/A, Budapest, Hungary
          }

\abstract{%
 Space photometric missions have been steadily accumulating observations of Cepheids in recent years, leading to a flow of new discoveries. In this short review we summarize the findings provided by the early missions such as \textit{WIRE, MOST}, and \textit{CoRoT}, and the recent results of the \textit{Kepler} and \textit{K2} missions. The surprising and fascinating results from the high-precision, quasi-continuous data include the detection of the amplitude increase of Polaris, and exquisite details about V1154 Cyg within the original \textit{Kepler} field of view. We also briefly discuss the current opportunities with the \textit{K2} mission, and the prospects of the \textit{TESS} space telescope regarding Cepheids.}

\maketitle

\section{Introduction}\label{sec:intro}
By instinct, Cepheid variables might not seem to be ideal targets for photometric space missions. Their variations are known to be rather simple: they pulsate in one or a few radial \textit{p}-modes in a rather clockwork-like manner. They often have long pulsation periods that may extend to several weeks. And we have excellent ground-based observations for many of them, particularly from the OGLE project, that made several interesting discoveries (see, e.g., \cite{ogle-blg,ogle-mc}). The expanding capabilities of time-domain astronomy provide us with ample data about Cepheid stars from the ground.

But is observing Cepheids from space really a moot point? We now have had missions like \textit{CoRoT} and \textit{Kepler} that observed long enough to cover several pulsation cycles of a typical Cepheid. Although long-term ground-based observations can reach very high precision for periodic signals, the per point accuracy of space telescopes is still unmatched. And as the results below illustrate, these benefits allowed us to uncover new, unexpected phenomena in Cepheids that have been not or are barely observable from the ground.  

Before we delve into the details, it is worth pointing out that the Hubble Space Telescope itself accumulated a large amount of photometry about Cepheids, mostly of extragalactic ones. However, the main goal of these observations was to pin down the period-luminosity relations for the various galaxies, not asteroseismology (for a summary of these observations, see \cite{riess}). Therefore we won't discuss these observations in detail here.

\section{Small(ish) space telescopes}
\subsection{Early efforts}
Space-based photometric telescopes were envisaged for a long time. The first instrument dedicated to variable stars was the French \textit{EVRIS} telescope that was flying---briefly---on the failed \textit{Mars-96} mission two decades ago \cite{evris}. Interestingly, the first actual photometric mission was born from another mishap: the small optical star tracker camera on the otherwise unusable \textit{WIRE} infrared space telescope was utilized to monitor bright stars. The well-known Cepheid Polaris ($\alpha$ UMi) was observed with it multiple times in 2004-5, complementing the more extended, but significantly lower-quality data set of the \textit{SMEI} all-sky camera aboard the \textit{Coriolis} satellite \cite{polaris}. These observations revealed that Polaris is not switching off and its pulsation amplitude started to increase after reaching a minimum around 2000. Weak excess variation in the low-frequency regime was tentatively connected to granulation patterns on the surface of the star \cite{polaris}. 

\subsection{The first photometric missions}
Meanwhile, missions dedicated to stellar variability were also launched into space. The Canadian \textit{MOST} and the French-led \textit{CoRoT} missions utilized larger telescopes, compared to the ones mentioned so far. While the former telescope is more or less limited to one month long observations, the latter was able to perform long runs lasting for 150 days. Unfortunately, almost all stars (a single type II and seven classical Cepheids) that fell into the observing fields of \textit{CoRoT} turned out to be very stable, with almost no signs of variations from one cycle to the next \cite{poretti-corot}. There was one star though, CoRoT 0223989566, that revealed an unexpected pulsation mode. This O1/O2 (first- and second-overtone) beat Cepheid also features a longer-period mode at $P_1/P_3 = 0.68$ ratio that has been also identified in a number of first-overtone RR Lyrae stars, possibly suggesting a common origin for the mode in the two classes \cite{poretti2014,netzel}.

The venerable \textit{MOST} space telescope observed four Cepheids during its lifetime, all of which belonged to different subgroups. First, it surveyed RT Aur and SZ Tau, a fundamental-mode and a first-overtone star, and looked for signs of irregularity in the pulsations. The results suggested that overtone stars may have less stable light curves \cite{evans-most}. Later, the telescope serendipitously observed the beat Cepheid U TrA, and carried out a targeted campaign on the modulated, second-overtone star V473 Lyr. The latter observations led to the first discovery of period doubling in a classical Cepheid, hinting that similar dynamical processes may occur in modulated Cepheid and RR Lyrae stars \cite{molnar-most}.

More recently, the BRITE-Constellation mission also covered some bright Cepheids. Although \textit{BRITE} utilizes multiple small satellites to observe the same field simultaneously in blue and red colors, Cepheids are too faint for them in the blue band. Red light curves have been obtained for seven targets so far, with varying success. Nevertheless, the observations revealed a possible modulation in T~Vul, and signs of additional modes in DT~Cyg and V1334~Cyg \cite{brite,britecep}.

\section{The \textit{Kepler} and K2 years}

\subsection{Four years with a Cepheid}
The field-of-view of the original \textit{Kepler} mission included a single Cepheid only, V1154~Cyg. The first months of data indicated that V1154~Cyg is a regular Cepheid pulsating only in the fundamental mode \cite{szabo2011}. More extended observations then revealed that the star is, in fact, not an entirely regular clock. Various measures of the shape, amplitude, and length of the pulsation cycles showed irregular jitter around the mean values that averaged out on longer time scales \cite{derekas2012}. 

And then the four-year-long data set revealed even more. A weak modulation cycle with a period of about 159 d was identified. But the length and precision of the data also made it possible to clearly detect granulation noise in a Cepheid for the first time. Interestingly, the star lacked any signs of solar-like oscillations well below the expected amplitudes scaled from red giants \cite{derekas2017}.

\subsection{Surveying Cepheids with K2}
The mechanical issues of the \textit{Kepler} space telescope led to the start of the K2 mission that opened up new possibilities to observe rare kinds of stars, including Cepheids. Multiple campaigns are aimed towards the vicinity of the galactic disk where most classical Cepheids reside. Campaigns 9 and 11, in particular, cover the bulge, and provide a unique opportunity to compare the long-term but sparse observations of OGLE with the short, but continuous data from K2. 

\begin{figure*}
\centering
\includegraphics[width=1.0\textwidth,clip]{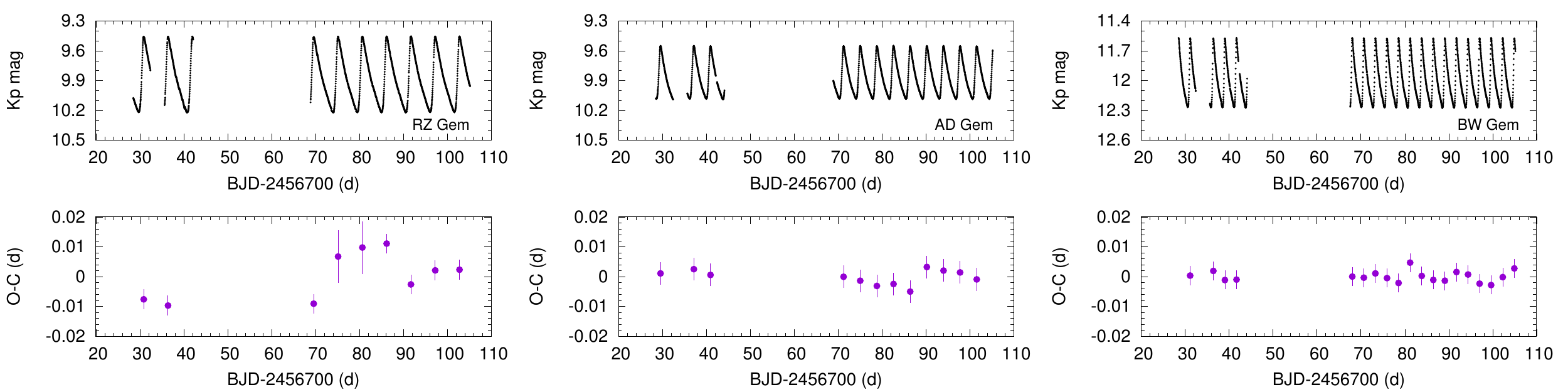}
\caption{Top: light curves of three fundamental-mode classical Cepheids observed in Campaign 0. Bottom: preliminary O-C values of times of maxima. }
\label{fig:fig-1}       
\end{figure*}

Cepheids were observed already in Campaign 0 towards the anticenter direction of the Galaxy. The field included three fundamental-mode classical Cepheids, BW~Gem, AD~Gem, and RZ~Gem. Since one of the discoveries of \textit{Kepler} about Cepheids was the detection of pulsation jitter, we examined the light variations of these three stars. Unfortunately, the light curve of the longest-period star, RZ~Gem suffers from jumps caused by the attitude changes of the telescope that need to be mitigated first. Nevertheless, we investigated the O--C values of the brightness maximum timings for all three stars. The scatter for RZ~Gem ($P$ = 5.528 d) was in the same range as that of V1154 Cyg ($P$ = 4.925 d), about $\pm$ 0.015 d. But as Fig.~\ref{fig:fig-1} shows, the scatter of maxima timings turned out to be much smaller ($\pm$ 0.005 d) for the two short-period stars, AD~Gem ($P$ = 3.788 d) and RZ~Gem ($P$ = 2.635 d). These first results hint that the amount of pulsation jitter might depend on the pulsation period of the star, but more detailed studies with larger sample sizes are required to draw more certain conclusions. Furthermore, observations of first-overtone and beat Cepheids will help us investigate the stability of different modes, as well as to search for low-amplitude additional modes.

The K2 mission also covers many type II and anomalous Cepheids that earlier missions largely lacked. The first results suggest that W~Vir-type stars (the medium-period type II subclass) may experience period doubling and more irregular cycle-to-cycle variations \cite{plachy2017}. The OGLE surveys already suggested that the pulsation of W Vir-type stars is not perfectly repetitive \cite{ogle-blg}, but now the K2 mission may show us how exactly the cycles vary. 

\begin{figure*}
\centering
\includegraphics[width=1.0\textwidth,clip]{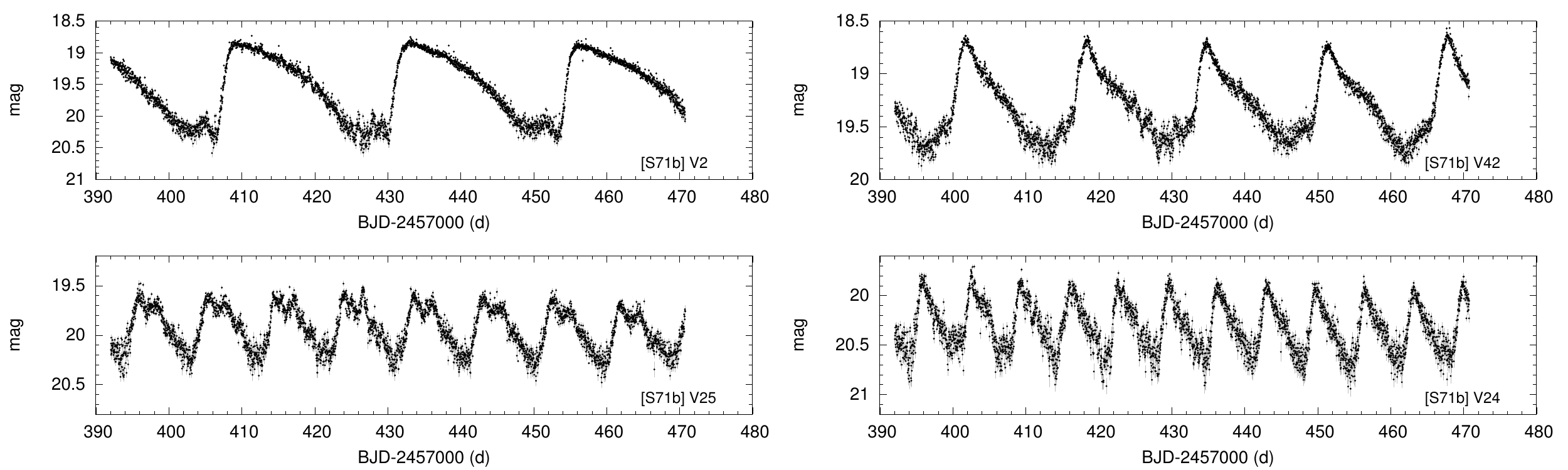}
\caption{Preliminary K2 light curves of various Cepheids in the Local Cluster galaxy IC 1613, obtained during Campaign 8.}
\label{fig:fig-2}       
\end{figure*}

Another unexpected capability of the K2 mission is that it is able to reach some extragalactic pulsating variable stars. First, it observed three RR Lyrae stars in the dwarf spheroidal galaxy Leo IV during Campaign 1 \cite{leoiv}. Then in Campaign 8 it observed a much larger irregular dwarf galaxy in the Local Cluster, IC 1613. While it is farther away than Leo IV, IC 1613 contains hundreds of Cepheids, among other bright variable stars, that \textit{Kepler} is capable to observe. The low angular resolution makes \textit{Kepler} a less than ideal tool to measure extragalactic variables that suffer from crowding and blending with nearby stars: nevertheless, extracting continuous light curves from even a fraction of the Cepheids in IC 1613 will be a unique result. At the time of writing the photometry is still under way, but we include preliminary light curves of a few bright targets here, in Fig.~\ref{fig:fig-2}, as an appetizer for the complete study. Photometry was carried out using the \textsc{fitsh} package, similar to the method used in case of Leo IV \cite{fitsh}. For IC 1613 we also included image oversampling into the pipeline to allow for subpixel-level image shifts that gives us a smoother master image for image subtraction. 

\section{Upcoming data sources}
More missions are under way: the European \textit{Gaia} mission will soon provide accurate geometric parallaxes for a large number of Cepheid stars, as well as sparse photometry for most of them, and potentially discover many, yet unknown members of the class \cite{gaiacu7}. The distance measurements of \textit{Gaia} will untangle many, currently uncertain parameters of these stars, including possible companion stars, the amount of infrared excess, and the exact values and possible period dependence of the radial-velocity projection factor \cite{kervella}. It will also allow us to properly separate the anomalous Cepheid population in the disk and halo from the long-period RR~Lyrae stars, based on their luminosities. 

The American \textit{TESS} space telescope will be a new source of continuous photometry from 2018 \cite{tess}. Unlike \textit{Kepler}, it will observe only the brighter stars in the sky, and, for the most part, in 27 d long runs. The short observing time per sector is not favorable for all but the shortest-period Cepheids. Fortunately, the continuous viewing zones (CVZ) of the mission offers one-year-long coverage. The CVZ towards the south ecliptic pole will enable us to observe the brightest, longest-period Cepheids in the Large Magellanic Cloud. The bright Galactic Cepheid, $\beta$~Dor is also included in the CVZ, and the data will potentially be accurate enough to look for the signs of granulation, and compare its properties with that of V1154~Cyg. And in the coming decade, the European \textit{PLATO} mission will expand these kinds of studies even further. 

However, single-color photometric observations will not tell us everything about these stars. Targeted efforts to measure accurate line profiles and radial velocities can gives us important clues about the physical processes in the atmospheres and the pulsation \cite{anderson,nardetto}. Multicolor photometry from large-\'etendue surveys also provide important physical information for us, and can be exploited for e.g., mode identification. And given the usually high apparent brightnesses of Cepheids, surveys that achieve the large \'etendue by combining small apertures with extremely large field of view (and high cadence), such as the Fly's Eye system \cite{flyseye}, are better suited at these tasks than larger telescopes such as PanSTARRS or LSST. The combination of such ground-based observations with space-based photometry may ensure that the golden age for stellar astrophysics continues.

\begin{acknowledgement} 
\noindent\vskip 0.2cm
\noindent {\em Acknowledgments}: L.~M. acknowledges with thanks the hospitality and financial support of the organizers of the 22nd Los Alamos Stellar Pulsation Conference Series meeting. L.~M. is supported by the J\'anos Bolyai Research Scholarship of the Hungarian Academy of Sciences. This project has been supported by the Lend\"ulet LP2012-31 and LP2014-17 Programs of the Hungarian Academy of Sciences, and by the NKFIH K-115709, PD-116175, and PD-121203 grants of the Hungarian National Research, Development and Innovation Office. This paper includes data collected by the Kepler and K2 missions, funded by the NASA Science Mission directorate. This work used K2 data proposed by WG7 of the Kepler Asteroseismic Science Consortium (GO numbers 0055 and 8041).
\end{acknowledgement}

%
%

\end{document}